\documentclass[journal=nalefd,manuscript=letter]{achemso}

\usepackage[version=3]{mhchem} 
\usepackage{graphicx}
\usepackage{dcolumn}
\usepackage{bm}
\usepackage{xr}
\usepackage{color}

\newcommand{\comment}[2]{#2}

\author{Sudipta Dubey}
\affiliation[]{Department of Condensed
Matter Physics and Materials Science, Tata Institute of Fundamental
Research, Homi Bhabha Road, Mumbai 400005, India}
\author{Vibhor Singh}
\affiliation[]{Department of Condensed
Matter Physics and Materials Science, Tata Institute of Fundamental
Research, Homi Bhabha Road, Mumbai 400005, India}
\author{Ajay K Bhat}
\affiliation[]{Department of Condensed
Matter Physics and Materials Science, Tata Institute of Fundamental
Research, Homi Bhabha Road, Mumbai 400005, India}
\author{Pritesh Parikh}
\affiliation[]{Department of Condensed
Matter Physics and Materials Science, Tata Institute of Fundamental
Research, Homi Bhabha Road, Mumbai 400005, India}
\altaffiliation{Birla Institute of Technology and Science, Pilani, Hyderabad, 500078, India}
\author{Sameer Grover}
\affiliation[]{Department of Condensed
Matter Physics and Materials Science, Tata Institute of Fundamental
Research, Homi Bhabha Road, Mumbai 400005, India}
\author{Rajdeep Sensarma}
\affiliation[]{Department of Theoretical Physics, Tata Institute of Fundamental
Research, Homi Bhabha Road, Mumbai 400005, India}
\author{Vikram Tripathi}
\affiliation[]{Department of Theoretical Physics, Tata Institute of Fundamental
Research, Homi Bhabha Road, Mumbai 400005, India}
\author{K. Sengupta}
\affiliation[]{Theoretical Physics Department, Indian Association for the Cultivation of Science, Kolkata 700032, India}
\author{Mandar M. Deshmukh}
\email{deshmukh@tifr.res.in}
\affiliation[]{Department of Condensed
Matter Physics and Materials Science, Tata Institute of Fundamental
Research, Homi Bhabha Road, Mumbai 400005, India}
\title
{Tunable Superlattice in Graphene To Control the Number of Dirac Points}


\begin{document}
\begin{abstract}

Superlattice in graphene generates extra Dirac points in the band structure and their number depends on the superlattice potential strength. Here, we have created a lateral superlattice in a graphene device with a tunable barrier height using a combination of two gates. In this Letter, we demonstrate the use of lateral superlattice to modify the band structure of graphene leading to the emergence of new Dirac cones. This controlled modification of the band structure persists upto 100 K.

\end{abstract}

\section{}

Chirality in graphene provides a unique platform for the experimental observation of phenomena like Klein tunneling\cite{klein1,shytov_klein_2008,klein2}, unusual integer quantum Hall effect\cite{novoselov_two-dimensional_2005,zhang_experimental_2005}, and rich fractional quantum Hall spectra\cite{bolotin_observation_2009,du_fractional_2009}. The carrier density in graphene can be changed by applying gate voltages, which allows one to study tunable plasmonic and photonic excitations\cite{chen_optical_2012_1,fei_gate-tuning_2012_1,yan_tunable_2012,koppens_graphene_2011,lee_vertical_2010}, Coulomb drag\cite{gorbachev_strong_2012} along with technological applications\cite{novoselov_roadmap_2012} like optical modulators\cite{liu_graphene-based_2011}, RF transistors\cite{wu_high-frequency_2011}, and ultrafast and high gain photodetectors\cite{xia_ultrafast_2009,konstantatos_hybrid_2012}. A technological challenge in working with graphene is the absence of a bandgap.  Several methods like chemical functionalization\cite{elias_control_2009}, nanoribbons\cite{li_chemically_2008}, uniaxial strain engineering\cite{pereira_tight-binding_2009} have been proposed and experimentally demonstrated to modify the band structure of graphene. However, ideas based on chemical functionalization, nanoribbons, and strain engineering do not provide any tunability of properties and their compatibility with large scale integration of devices is yet to be tested. Esaki et al. \cite{tsu_tunneling_1973} first proposed the use of a periodic superlattice (SL) potential to modify the band structure in semiconductors and similar experiments have been proposed in monolayer and bilayer graphene\cite{park_anisotropic_2008,Diracptsquarepot,killi_band_2011,burset_transport_2011,killi_graphene:_2012,young_electronic_2011,tan_new_2011}. In contrast to conventional materials, SL in graphene results in anisotropic renormalization of the velocities of the Dirac quasiparticles\cite{Diracptsquarepot,park_new_2008} leading to the possibility of collimation\cite{park_electron_2008}. It also generates extra Dirac points\cite{park_anisotropic_2008,Diracptsquarepot,extradirac} in the band structure and these have been experimentally observed in graphene where the substrate induces a periodic potential\cite{boronnitride,ponomarenko_changes_2012,dean_hofstadters_2012,pletikosic_dirac_2009}. Superlattices using Moire pattern\cite{carozo_raman_2011} by laying graphene over hexagonal boron nitride substrates\cite{boronnitride} have been demonstrated, but they suffer from the lack of tunability of the superlattice potential. For the first time, here we have created a lateral SL in a graphene device with a tunable barrier height using a combination of two gates with the top-gate consisting of a comb like structure pinned to the same potential. An ability to engineer the band structure of graphene using SL, as we demonstrate, opens up new possibilities.

\begin{figure}
\includegraphics[width=150mm]{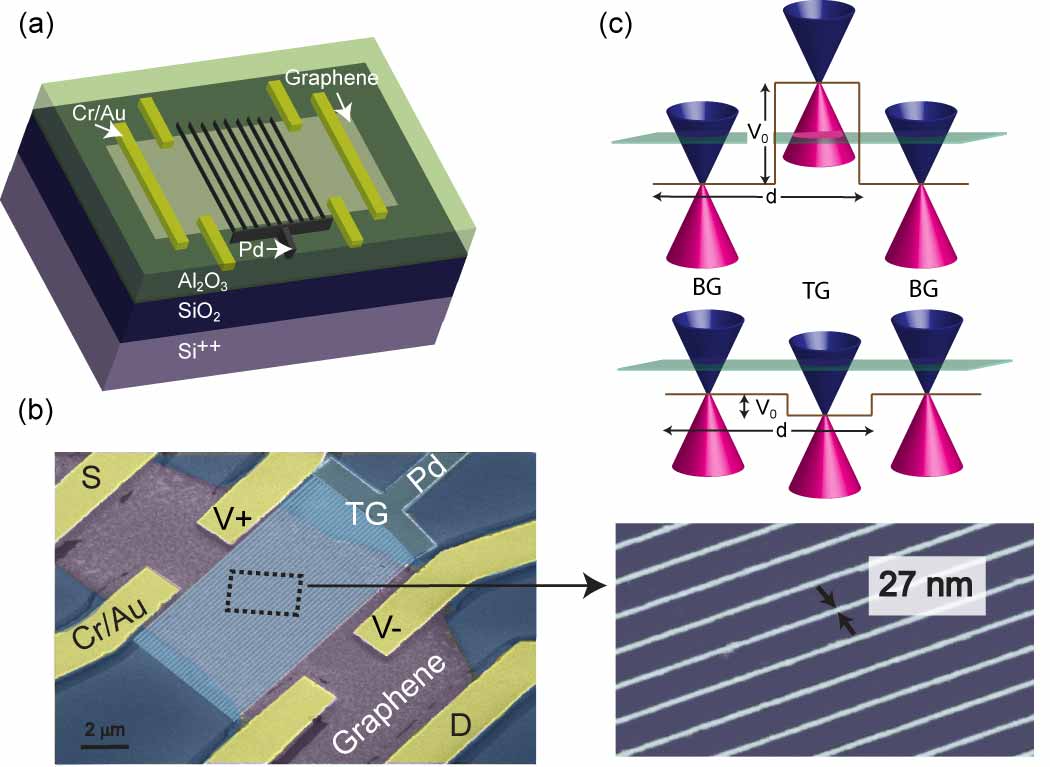}
\caption{\label{fig:figure1} Device schematic to realize a lateral graphene SL. \comment{(c) Potential barriers created by alternate regions of local doping on a planar graphene device; this is realized using a back-gate and a series of top-gates pinned to the same potential.} (a) Schematic of the device with a back-gate and a series of top-gates pinned to the same potential. This helps realize a regular periodic potential with $\sim$ 40 \comment{40 \textbf{check}} periods. The substrate for the device is 300~nm thick silicon dioxide on degenerately doped silicon. The graphene flake is contacted by depositing \comment{\textbf{check}}10~nm of Cr and \comment{\textbf{check}20}50~nm of Au. The top-gate dielectric is \comment{\textbf{check}20}23~nm of Al$_2$O$_3$ with 3~nm deposited by e-beam evaporation and 20~nm deposited using ALD. The top-gates are defined by electron beam lithography and by evaporating Pd. The width of top-gates is $\sim$ 27~nm with a period of 150~nm. \comment{and the spacing between adjoining top-gates is 125~nm. Due to the finite thickness of the top-gate dielectric, \comment{the potential in the graphene flake has a total period of 150~nm with} effective width of the top-gate region in graphene flake \comment{with potential to be} $\sim$ 60-70~nm.} \comment{75~nm}(b) False colored SEM image of a device in Hall bar geometry showing the device dimensions. The arrow points to a \comment{inset of the top-gate region showing a} magnified view of the fingers of the top-gate structure. (c) Potential barriers created by alternate regions of local doping on a planar graphene device; this is realized using a back-gate and a series of top-gates pinned to the same potential. In the top figure, TG region is hole doped and BG region is electron doped and in the bottom figure both TG and BG region are electron doped with the green plane showing the Fermi energy. }
\end{figure}

To achieve this goal we fabricate devices in Hall bar geometry with multiple thin top gates defined over the central region of graphene as shown in the schematic in Figure~\ref{fig:figure1}a. \comment{Our device geometry is illustrated in Figure 1(a) and.} Figure~1b shows a scanning electron microscope (SEM) image of a device showing multiple ($\sim$27~nm wide) Pd top gates with a period of 150~nm (details in Section~1 of Supporting Information). The graphene flake can be divided into a series of alternating regions, one with only a back-gate (region denoted as BG)  and the other that has both a top-gate (tg) and a back-gate (bg) (region denoted as TG) (see Figure~1c). The charge density in BG region is determined only by the applied back-gate voltage, while in the TG region, it is determined by both the back-gate voltage $V_{bg}$ and the top-gate voltage $V_{tg}$, providing an independent control to set the charge carrier density and type in the two regions. The amplitude of the SL potential created is the difference in the charge neutral point between the BG and the TG region (illustrated in Figure~1c). The capacitive coupling of the top-gate ($C_{tg}$) is higher than that of the back-gate ($C_{bg}$) due to the geometrical proximity and higher dielectric constant of top-gate dielectric ($C_{tg}/C_{bg} =$ 18) \comment{, where $C_{tg}$ and $C_{bg}$ are capacitance per unit area of the graphene sheet to top-gate and back-gate respectively (details regarding capacitance ratio in)} (see Section~4 of Supporting Information for details). Fringing fields from the top-gate electrodes lead to the smoothening of the SL potential and the effective width of the top-gate felt by the charge carriers in graphene is approximately 60-70 nm (details regarding the potential profile in Section~9 of the Supporting Information). \comment{The period of top-gates is designed to be 150~nm.} Varying $V_{tg}$ and $V_{bg}$, we probe the system having a series of \emph{p-n'-p} (or \emph{n-p'-n}) or \emph{p-p'-p} (or \emph{n-n'-n}) junctions depending on the combination of the two gate voltages, giving rise to a SL structure in graphene. The advantage of creating a SL structure using gate voltage is that the amplitude of the SL potential can be continuously varied, giving one control over the electronic properties, which is desirable for device applications. In addition, studies have suggested that around 10 periods of the SL are sufficient to see the effect of band formation \cite{brey_emerging_2009}; in our experiments we have used $\sim$ 40 periods of the SL.

\begin{figure}
\includegraphics[width=170mm]{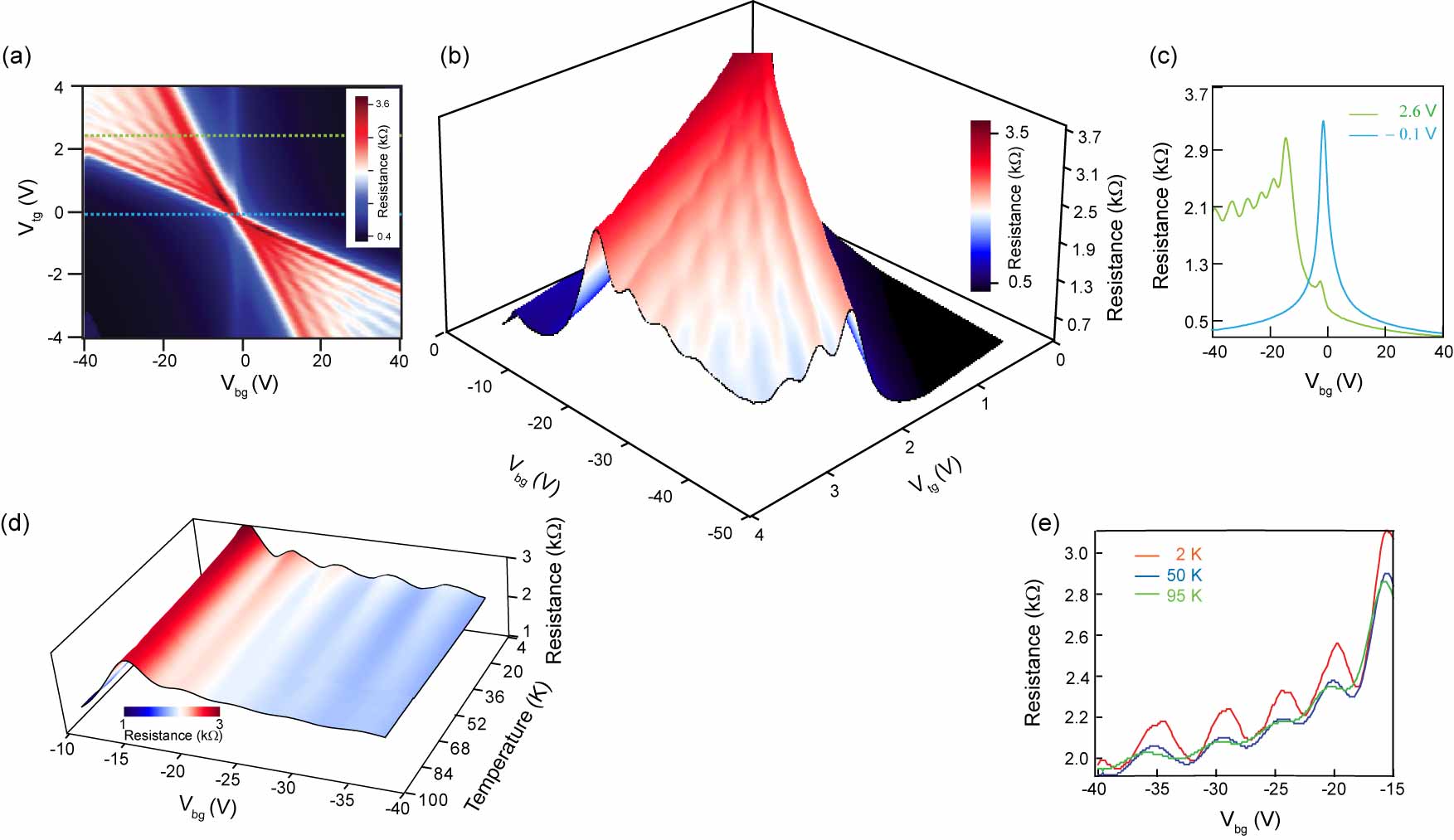}
\caption{\label{fig:figure2} Electron transport measurements to probe the properties of the SL structure. (a) Resistance measured as a function of back-gate voltage $V_{bg}$ and top-gate voltage $V_{tg}$ at $300$ mK. The oscillations in the resistance are seen when charge carriers in the TG and BG regions are of opposite sign. \comment{ induced by the back-gate and in the region under top-gate are of opposite sign.} A positive (negative) sign of the gate voltage induces electrons (holes) in the system. The charge density in TG region \comment{between a top and bottom gate are} is set by the algebraic sum of the effect of the voltages applied to the two gates. (b) A detailed measurement of the region of one of the quadrants shown in (a) where $V_{bg}$ induces holes in the graphene sheet and $V_{tg}$ induces electrons giving rise to a potential barrier of varying height at different points in the region. The splitting of resistance peaks as a function of the gate voltages is clearly observed. \comment{The dashed lines indicate the evolution of peaks and their splitting.} (c) Line plot of data shown in (a) at $V_{tg}$ = -0.1~V (blue) and $V_{tg} =$ 2.6~V (green). The line plot for $V_{tg} =$ 2.6~V depicts oscillations. \comment{The arrow indicates the charge neutrality in the region which is not covered by top-gate.} (d) Resistance as a function of $V_{bg}$ and temperature with  $V_{tg} =$ 2.6~V showing that the oscillations persists even at around 70~K. (e) Line plot of resistance as a function of $V_{bg}$ with  $V_{tg} =$ 2.6~V at three different temperatures. \comment{(f) Variation of the amplitude of the conductance modulation as a function of temperature. (f) Resistance as a function of $V_{bg}$ and magnetic field when $V_{tg} =$ 2.6~V. The arrows show the phase shift of the oscillations with magnetic field signifying Klein tunneling through the SL barriers \cite{shytov_klein_2008}. (g) Line plot of resistance as a function of $V_{bg}$ ($V_{tg} =$ 2.6~V) at two different magnetic field of 0~T and 1~T showing the change from peak to dip in the oscillations (e) Resistance as a function of $V_{bg}$ and temperature with  $V_{tg} =$ 2.6 V. (f) Resistance as a function of $V_{bg}$ with  $V_{tg} =$ 2.6~V as a function of temperature at three different temperatures. The inset shows how the amplitude of the resistance oscillation changes as a function of temperature.}}
\end{figure}

The electrical properties of a  monolayer graphene device (evidence for monolayer graphene provided in Section~2 of Supporting Information) \comment{Section~II)} are studied using a zero bias measurement. Figure~\ref{fig:figure2}a shows the colorscale plot of resistance measured at~300~mK as a function of $V_{bg}$ and $V_{tg}$. The resistance of the device is low when charge carriers are of the same type in TG and BG region ($sgn(V_{bg}V_{tg})$=1) (top right and bottom left quadrant in Figure~\ref{fig:figure2}a). However, when the type of charge carriers in the two regions are different (this happens in part of the bottom right and top left quadrant of Figure~\ref{fig:figure2}a), we observe oscillations in resistance as a function of $V_{tg}$ and $V_{bg}$. These oscillations fade in and out; their number increases with increasing gate voltage as seen \comment{illustrated by the dashed black lines} in Figure~\ref{fig:figure2}b, which is a magnified and detailed measurement of one quadrant of Figure~\ref{fig:figure2}a. \comment{The resistance peaks are indicated by dashed lines and} The number of resistance ``ridges" increases by one as the magnitude of $V_{tg}$ and $V_{bg}$ is gradually increased. We note that increasing $V_{bg}$ and $V_{tg}$ in these quadrants amounts to increasing the superlattice barrier seen by the charge carrier and we discuss this aspect later in greater detail. Fixing $V_{tg}$ at 2.6~V, resistance as a function of $V_{bg}$ shows a distinct oscillatory pattern (as shown in green curve in Figure~\ref{fig:figure2}c). A slice of the data at the charge neutrality point shows a peak in the resistance at $V_{bg}=$ -2~V as seen in the blue curve of Figure~\ref{fig:figure2}c. The field effect mobility of this device is $\sim$ 6000~cm$^2$/(Vs) at 300~mK and corresponds to a mean free path of $\sim$ 70~nm (details regarding mobility calculation provided in Section~3 of Supporting Information). \comment{Section~III).} The presence of the array of top-gates does not significantly affect the mobility of the device (details in Section~3 of Supporting Information). Another aspect relevant for the transport properties of graphene devices is the depth of the charge puddles \cite{martin_observation_2008,deshpande_spatially_2009,rossi_ground_2008,dean_boron_2010}, U$_{pud}$, that result from the inhomogeneity of charge distribution. Here, U$_{pud}~\sim$~71~meV (see Section~3 of Supporting Information) \comment{Section~III)} and is comparable with that seen by others in devices with similar mobility.

To further probe the energy and length scale related to the resistance oscillations, we study the transport as a function of temperature. Figure~\ref{fig:figure2}d shows the measurement of resistance with gate voltage and temperature (up to 100 K). The oscillations in resistance with gate voltage are robust against temperature and can be seen upto 100~K (8.3~meV), which provides an estimate of the relevant energy scale for the oscillations. \comment{ as shown in Figure~\ref{fig:figure2}(d), where we have varied the temperature upto 100~K.} The amplitude of the oscillations decrease with increasing temperature, as is clearly seen in the line plots taken at three different temperatures (Figure~\ref{fig:figure2}e).(Additional data on temperature variation of oscillations is shown in Section~6 of Supporting Information.) \comment{{\color{red}This sets an upper bound on the energy involved in the process responsible for the oscillations.}} \comment{and the trend in the variation of the amplitude of oscillations at different $V_{bg}$ is seen in Figure~\ref{fig:figure2}(f).}


We would like to note that the resistance oscillations are not as a result of coherent Fabry-Perot resonance of the SL, since the phase coherence $l_\phi\sim$ ~0.6$~\mu$m at 300 mK (details provided in Section~5 of Supporting Information), whereas the SL periodicity is 150~nm. Therefore phase information is lost after 4 SL periods whereas in transport measurement we probe 40 SL periods. The fact that the oscillations persist upto 100~K, where $l_\phi$ is very small, indicates that the band picture of SL rather than the coherent Fabry-Perot resonances is \comment{required} meaningful in understanding the oscillations in our experiment.

Having considered the experimental results, \comment{from our devices} we now try to understand the potential profile created due to the combination of gates. To the first approximation, we assume that the potential created due to the top-gate is abrupt. The height of the potential created due to the top-gate and back-gate can be calculated from the doping in the two regions, with and without the top-gate; this amounts to measuring the shift in the charge neutrality point in the two regions\cite{klein1}. The charge density induced by the back-gate is $C_{bg}V_{bg}$ and that by the top-gate is $C_{tg}V_{tg}$. The amplitude of the SL potential barrier $V_0$ \comment{, created by both $V_{tg}$ and $V_{bg}$} is then given by (details in Section~8 of Supporting Information)
\begin{equation}\label{eq:eq1}
V_0 = \sqrt{\pi}\hbar v_F \left(sgn(C_{bg}V_{bg})\sqrt{\frac{|C_{bg}V_{bg}|}{e}} -sgn(C_{tg}V_{tg}+C_{bg}V_{bg})\sqrt{\frac{|C_{tg}V_{tg}+C_{bg}V_{bg}|}{e}}\right)
\end{equation}
It is to be noted that for constant amplitude of the potential, in the bottom right quadrant of Figure~\ref{fig:figure2}a, we have electrons coming across barriers of constant height, and in the top left quadrant, holes face wells of constant depth. Since the measured data is symmetric, henceforth the magnitude of SL barrier (or well) is referred to as $V_0$.

\begin{figure}
\includegraphics[width=150mm]{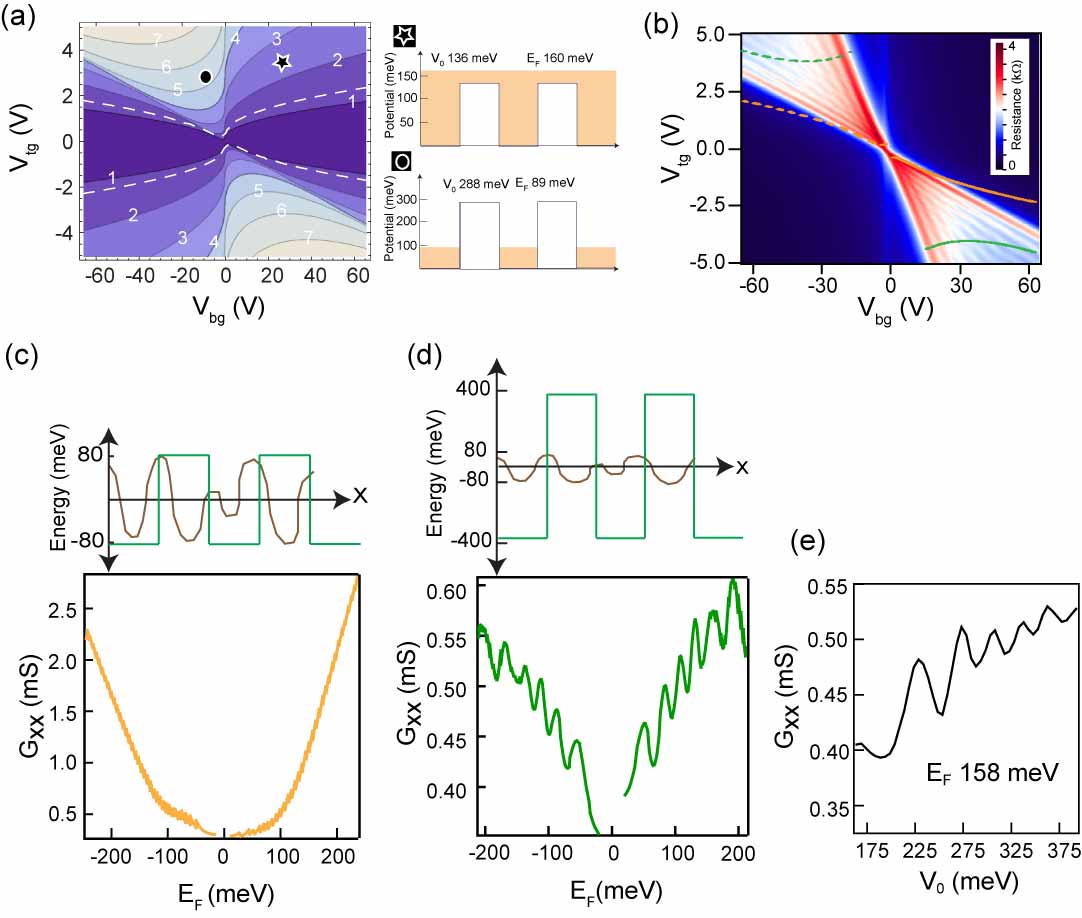}
\caption{\label{fig:figure3}  Effect of periodic potential on the transport. (a) Calculated height of potential barrier $(V_0/E_{SL})$ \comment{$u = V_0d/\hbar v_F$} as a function of $V_{tg}$ and $V_{bg}$ where $V_0$ is given by eq~\ref{eq:eq1}. The plot shows contours of $\frac{V_0}{E_{SL}}$ in units of $4 \pi$. The white dashed contour correspond to U$_{pud}$, and we observe effect of SL only when $V_0>>$U$_{pud}$. It is clear that the strength of the potential increases as the top-gate and back-gate strength increases to induce carriers of opposite sign (with  $sgn(V_{tg}V_{bg})$=-1). Schematic on the side illustrates the height of the SL potential and the position of Fermi energy at two different regions denoted by star (region where oscillations are not observed) and circle (region where we see oscillations). (b) Experimentally measured resistance as a function of $V_{tg}$ and $V_{bg}$ with curves of constant $V_0$.
Contour of constant potential $(V_0/E_{SL}) = 6\pi$ (in yellow) and $(V_0/E_{SL}) = 26\pi$ (in green) overlaid on resistance as a function of $V_{bg}$ and $V_{tg}$ at 300~mK. This allows us to extract from experimentally measured data resistance variation at constant $V_0$ as a function of $V_{bg}$ and consequently $E_F$. (c) Conductance as a function of $E_F$ for $(V_0/E_{SL}) = 6\pi$. (d) Conductance as a function of $E_F$ for $(V_0/E_{SL}) = 26\pi$. In this case, we see oscillations clearly as the potential created by the gate voltage is much greater than the typical potential variations due to electron hole puddles U$_{pud}$. (see Section 8 of Supporting Information for slices at various $V_0$.) (e) Conductance as a function of $V_0$ at constant Fermi energy shows oscillatory behavior.}
\end{figure}

Introduction of the SL potential of period $d$ introduces another energy scale $E_{SL}=(\hbar v_F/d)$ which controls the effect of the SL on the system and we examine the experimental results in units of $E_{SL}$. For our experiments, $d =$ 150~nm and hence $E_{SL}$ is $\sim$~4.4~meV. Following eq~\ref{eq:eq1}, Figure~\ref{fig:figure3}a shows a contour plot of $V_0$ as a function of $V_{tg}$ and $V_{bg}$ for the range of parameters used in our experiments with contours at \comment{ . The contours are for} $(V_0/E_{SL})=4 \pi j$, where $j$ is an integer. \comment{Following Equation~\ref{eq:eq1}, Figure~\ref{fig:figure3}(a) shows the plot of $V_0$ as a function of both $V_{bg}$ and $V_{tg}$, and consequently,} To understand the experimental results, we have taken slices of the experimental data along contours of constant $V_0$. We take slices from the charge neutrality point and note that the charge neutrality point is influenced to some extent by both the gates due to fringing of fields resulting from the structure of the gates. \comment{(details of finite element method (FEM) calculation to model potential profile in Section~VIII of Supporting Information).} Figure~\ref{fig:figure3}b shows the contours along which the slices are taken (dashed for holes corresponding to negative $E_F$) (additional slices are shown in  Section~8 of Supporting Information). \comment{in Section~VI).}


\comment{We note that as seen in Figure~\ref{fig:figure3}(a), $V_0$ is a function of both $V_{bg}$ and $V_{tg}$, and consequently, to understand the experimental results, we have taken slices of the experimental data along contour of constant $V_0$. We take slices from the charge neutrality point and note that the charge neutrality point is influenced to some extent by both the gates due to fringing of fields due to the structure of the gates. (details of finite element method (FEM) calculation to model potential profile in Section~VIII of Supporting Information). Figure~\ref{fig:figure3}(b) shows the contours along which the slices are taken (dashed for holes corresponding to negative $E_F$) (additional slices are shown in Supporting Information in Section~VI).} Figure~\ref{fig:figure3}c,d shows slices of the data depicting longitudinal conductance (G$_{xx}$) for  $(V_0/E_{SL})=$ 6$\pi$ and 26$\pi$ respectively. It can be seen from Figure~\ref{fig:figure3}c that for $(V_0/E_{SL})=$ 6$\pi$, we do not observe oscillations in G$_{xx}$ as $E_F$ is increased. However for higher $V_0$, oscillations in G$_{xx}$ become clearly visible as $E_F$ is swept; \comment{$\frac{V_0}{E_{SL}}$ we observe clear oscillations in G$_{xx}$ as $E_F$ is swept;} an example of this is seen in  Figure~\ref{fig:figure3}d for $(V_0/E_{SL})=$ 26$\pi$. \comment{This sets a threshold on the onset of oscillations in conductance.} The reason for this observation is the inhomogeneous charge distribution with U$_{pud}\sim$ 71~meV (details regarding estimation about electron-hole puddles in graphene in Section~3 of Supporting Information). Because of the presence of puddles, the effects of the periodicity of the SL is not observed unless the depth of the potential modulation $V_0$ is much larger than U$_{pud}$; this is illustrated by the cartoons in Figure ~\ref{fig:figure3}c,d. \comment{ (see Section~VI of the Supporting Information for variation of $G_{xx}$ as a function of $V_0$).} Conductance as a function of $V_0$ at constant $E_F$ also shows oscillatory behavior as seen in Figure ~\ref{fig:figure3}e.

\comment{Another aspect where} The relevance of $E_{SL}$ is also observed in the temperature evolution of the resistance oscillations as seen in Figure~\ref{fig:figure2}d. For our experiment, $E_{SL}$ $\sim$~4.4~meV, and we observe that the oscillations in resistance, both as a function of $V_{tg}$ and $V_{bg}$, become indistinct around 100~K $\approx$~8.3~meV. The observation that the resistance oscillations are suppressed when temperature exceeds $E_{SL}$ and are seen only when $V_0 >>$ U$_{pud}$ shows that the resistance oscillations arise from the effect of SL potential and this serves as an internal consistency check. 

\begin{figure}
\includegraphics[width=150mm]{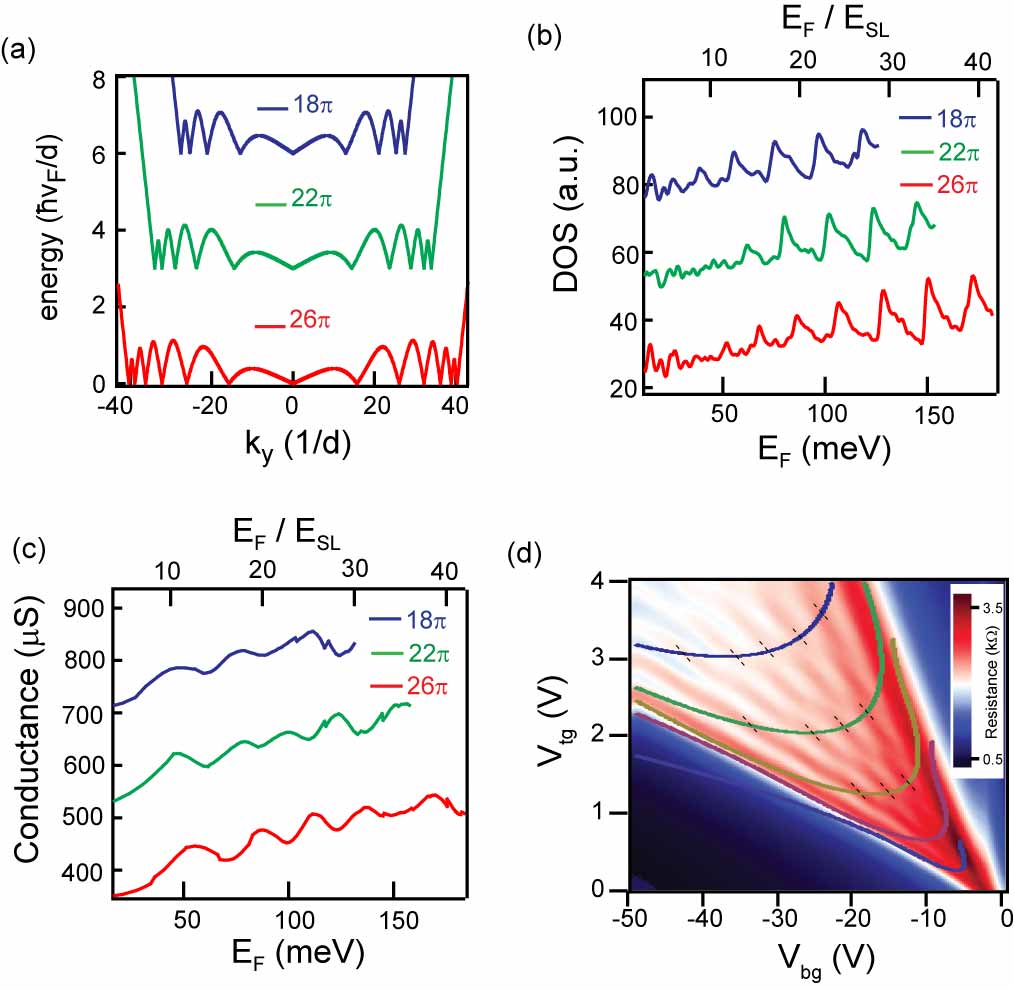}
\caption{\label{fig:figure4} Modeling the transport through the SL and comparison with experimental results. (a) Calculated dispersion relation of the conduction band at k$_x=$0 ($x$-direction is along the length of graphene and $y$-direction is along the width of graphene) for $V_0/E_{SL}=$18$\pi$, 22$\pi$, and 26$\pi$. As SL potential increases the band bends increasing the number of Dirac points. The plots are offset by 3 units for clarity. (b) Calculated DOS as a function of $E_F$ for the same strengths of potential $(V_0/E_{SL})$ as in (a). The plots are offset for clarity. DOS shows oscillations with an increasing trend with $E_F$ and the number of oscillations increase with increasing SL potential. (c) Plot of measured conductance as a function of Fermi energy for same value of $(V_0/E_{SL})$ as in (a,b). The plot for $(V_0/E_{SL})$ at 22$\pi$ and 18$\pi$ is shifted by 200~$\mu$S and 400~$\mu$S, respectively, for clarity. Similar to the oscillations observed in (a), we see an increasing trend in oscillations with $E_F$ and that their number increases with increasing SL potential. (d) Contours of constant $V_0$ separated by 4$\pi E_{SL}$ overlaid on measured resistance as a function of $V_{bg}$ and $V_{tg}$. Contours are calculated using the model of square SL. These contours pass approximately through the local peaks (shown by the black dotted line) showing that the number of oscillations increases by one as we cross the region of $V_0$ where one more extra Dirac point is created. We also observe that in the region of $j$ extra Dirac points, the number of oscillations is $j$, $j$ being a positive integer.}
\end{figure}

We have already seen indirect experimental evidence (in the form of the energy scale set by the temperature dependence of the oscillations) that the presence of the SL is related to the resistance oscillations. We will now discuss how the SL modifies the graphene band structure, which leads to the experimentally observed resistance oscillations. Several approaches have been followed to study the effects of SL in monolayer and bilayer graphene \cite{park_anisotropic_2008,Diracptsquarepot,killi_band_2011,burset_transport_2011,brey_emerging_2009}. Here, we will follow the theoretical approach of Barbier \emph{et al.}\cite{Diracptsquarepot}, which obtains the band dispersion of Dirac particles in a square SL potential (with periodic variation along the length of the sample and a constant profile along the width) using a transfer matrix method. We would like to note that, although we use this specific method for ease of application, other approaches also provide very similar predictions \cite{park_anisotropic_2008,Diracptsquarepot,killi_band_2011,burset_transport_2011,brey_emerging_2009}. The band dispersion of the conduction band of the above mentioned Kronig Penny model for Dirac fermions is plotted as a function of the momentum in the $y$ direction, $k_y$, for $k_x=$0 in Figure~\ref{fig:figure4}a. From top to bottom, the three curves correspond to $V_0/E_{SL}=$18$\pi$, 22$\pi$, and 26$\pi$, respectively. The most dramatic modification of the graphene band structure brought about by the SL  is the appearance of additional Dirac points (other than the one at $k_y=$0), as seen in this figure. The number of additional Dirac points increase by 1 as $V_0/E_{SL}$ is increased by 4$\pi$ (e.g., from 18$\pi$ which has four additional  Dirac points to 22$\pi$, which has five additional Dirac points).

The relation between appearance of the Dirac points and the resistance oscillations can be understood by looking at the single particle density of states (DOS) calculated from the band dispersion, which is plotted in Figure~\ref{fig:figure4}b. The DOS oscillates with the energy (over and above a linearly increasing trend) with peaks corresponding to van Hove singularities occurring between the Dirac points. The number of these peaks follows the number of additional Dirac points in the spectrum and increase with increasing $V_0$.

The DOS is related to the conductivity through the Einstein relation, $\sigma=e^2\nu(E_F)D$, where $D$ is the diffusion constant and $\nu(E_F)$ is the DOS at the Fermi level. In our samples, which have low mobility and hence are in the low diffusive limit where the localization length is much larger than the inter-carrier separation (see Section 10 of Supporting Information for details), the electrical conductivity is dominated by inelastic scattering processes, which allows hopping of the electrons from their localized (but overlapping) wavefunctions. In this case, on dimensional grounds, the diffusion constant $D\sim \alpha \xi_l^2/\tau_{in}$, where $\alpha$ is a dimensionless constant, $\xi_l \gg k_F^{-1}$ is the localization length of the electrons and $\tau_{in}$ is the characteristic inelastic scattering time\cite{thouless_effect_1980}. In this limit, the conductivity simply mimics the DOS at Fermi level. The resistance oscillations we observe are thus a reflection of the oscillation of the DOS with energy with the number of oscillations tracking the number of additional Dirac points produced in the spectrum by the SL.

In Figure~\ref{fig:figure4}c, we plot the experimentally measured conductance as a function of the Fermi energy of the charge carriers. We plot the theoretical data for DOS in Figure~\ref{fig:figure4}b and the experimental data for conductance in Figure~\ref{fig:figure4}c for the same values of $V_0/E_{SL}$ (with the contours of constant $V_0$ shown in the $V_{bg}$-$V_{tg}$ plane in Figure~\ref{fig:figure4}d) and see that they compare well with each other.

Qualitatively, both show a number of oscillations on top of a linearly increasing trend with the number increasing with increasing SL potential. We now proceed to a more quantitative comparison of the theoretical predictions and our experimental data, focusing chiefly on the period of oscillations observed in the two cases. The theoretical DOS oscillates with a period
of $\sim$ 21.1~meV, and the experimentally measured conductance oscillates with a period of $\sim$ 27.9~meV. We would like to point out one major factor which can account for this difference, namely, the much smoother profile of the experimental superlattice potential compared to the abrupt square waveform of the Kronig Penny model used in the theory. Brey \emph{et al.}\cite{brey_emerging_2009} have studied the appearance of extra Dirac point with a smooth potential profile given by $V_0\cos(2\pi x/d)$ and have found that the condition for appearance of new Dirac peaks is given by $J_0(V_0d/2\pi\hbar v_F )=0$. The successive Dirac peaks in this case appear  when the SL potential is increased by $\sim 6.2\pi E_{SL}=1.55(4\pi E_{SL})$. Thus the period can be significantly affected due to the smoothness of the potential profile. Because of finite thickness of the top gate dielectric, our device lies somewhere in between these two extreme limits. Thus the smoothness of the potential profile qualitatively explains the difference observed between the experimental observations and the theoretical predictions from the simple model. For a more accurate match of theory and experiment, one would have to solve for the dispersion using the actual potential profile of the device, but such calculation would not provide any new insight into the universal property of such devices.

\comment{Figure~\ref{fig:figure4}(c) is a plot of calculated dispersion of the conduction band at k$_x=0$ summarizing the key point that as the SL potential increases, band structure in graphene gets modified, increasing the number of Dirac points by one.} Figure~\ref{fig:figure4}d summarizes the main experimental observation showing that as $(V_0/E_{SL})$ increases by $\sim$ 4$\pi$, that is, as the number of extra Dirac points increases by one, the number of oscillations increases by one. Another remarkable feature in the data seen in the plot is that in the region of $j$ extra Dirac points, the number of oscillations is $j$; $j$ being a positive integer.

In this work we have demonstrated a tunable SL resulting in controllable modification of the band structure. Besides electronic properties, such devices may be of interest for plasmonics and magnetic superlattices could be of interest for spintronic applications. SL structure modifying the band structure also opens ways to realize Weyl semimetal\cite{burkov_weyl_2011} predicted in topological insulators and for thermoelectric and spintronic applications\cite{song_interfacial_2010}. Our work is a step towards exploring such devices.



\comment{\textbf{ summary, mention problem open questions, suggestions: Although, our experiment does not independently test the collimation of charge carriers but indirectly the appearance of resistance oscillations  }}

\begin{acknowledgement}

We thank Professor Jim Eisenstein for helpful discussions, Dr. Abhilash T.S. for help with the measurement setup and Tanuj Prakash for help with fabrication. We thank Dr. Michael Barbier for sharing the scheme of their calculation. We acknowledge Swarnajayanthi Fellowship of Department of Science and Technology and Department of Atomic Energy of Government of India for support.

\end{acknowledgement}

\comment{
\begin{suppinfo}
Additional information and figures.
\end{suppinfo}
}

\providecommand*\mcitethebibliography{\thebibliography}
\csname @ifundefined\endcsname{endmcitethebibliography}
  {\let\endmcitethebibliography\endthebibliography}{}

\comment{
\begin{tocentry}
\includegraphics[scale=0.25]{TOCgraphic.jpg}\label{fig:toc}
\end{tocentry}
}
\end{document}